\documentclass[superscriptaddress,aps,prl,reprint,floatfix]{revtex4-1}
\usepackage{graphicx}% Include figure files
\usepackage{dcolumn}% Align table columns on decimal point
\usepackage{bm}% bold math
\usepackage{mathrsfs}
\usepackage{etex}
\usepackage{amsmath}
\usepackage{amssymb}
\usepackage{amsthm}
\usepackage{amsbsy}
\usepackage{breqn}
\usepackage[colorinlistoftodos]{todonotes}
\usepackage[colorlinks=true, allcolors=blue]{hyperref}
\usepackage{bm}
\usepackage{tabularx}
\usepackage{mathtools}
\usepackage{cleveref}
\usepackage{listings}
\usepackage{verbatim}
\usepackage{tikz}
\usepackage{multirow}
\usepackage{rotating}
\usetikzlibrary{positioning}
\usepackage{color}
\newcommand{\beq}{\begin{equation}}
\newcommand{\eeq}{\end{equation}}
\newcommand{\beqa}{\begin{eqnarray}}
\newcommand{\eeqa}{\end{eqnarray}}
\newcommand{\bem}{\begin{math}}
\newcommand{\eem}{\end{math}}
\newcommand{\rar}{{\rightarrow}}

\begin{document}

%\preprint{APS/123-QED}

\title{Steady state running rate sets the speed and accuracy of accumulation of swimming bacterial populations
}% Force line breaks with 

\author{Margaritis Voliotis}
\affiliation{College of Engineering$,$ Mathematics and Physical Sciences$,$ University of Exeter$,$  Exeter EX4 4QF$,$ UK}
\email[]{M.Voliotis@exeter.ac.uk}
\author{Jerko Rosko}
\affiliation{Laboratoire Jean Perrin$,$ Sorbonne Universit\'e$,$ Paris 75005$,$ France}
\author{Teuta Pilizota}
\affiliation{Centre for Synthetic and Systems Biology$,$ University of Edinburgh$,$ Edinburgh$,$ UK}
\author{Tanniemola Liverpool}
\affiliation{School of Mathematics$,$ University of Bristol$,$ Fry Building$,$  Bristol  BS8 1UG$,$ UK}
\affiliation{BrisSynBio$,$ Life Sciences Building$,$ University of Bristol$,$  Bristol  BS8 1TQ$,$ UK}
\email[]{t.liverpool@bristol.ac.uk}
\date{\today}
\begin{abstract}
We study the chemotaxis of a population of genetically identical swimming bacteria undergoing run and tumble dynamics driven by stochastic switching between clockwise and counterclockwise rotation of the flagellar rotary system. 
Understanding chemotaxis quantitatively requires that one links the switching rate of the rotary system in a gradient of chemoattractant/repellant to experimental measures of 
the efficiency of a population of bacteria in moving up/down the gradient. Here we achieve this by using a probabilistic model and show that the response of a population to the gradient is complex. 
We find the changes to a phenotype (the steady state switching rate in the absence of gradients) affects the average speed of the response as well as the width of the distribution and both must be taken into account to optimise the overall response of the population in complex environments.
%Using sophisticated analyses of rare events (large deviations from the mean) in the stochastic dynamics, 
This is due to the behaviour of individuals in the 'tails' of the distribution. Hence we show that for chemotaxis, the behaviour of atypical individuals can have a significant impact on the fitness of a population.
\end{abstract}
\maketitle

%\tableofcontents
Bacterial self propulsion, in particular flagellated motility, is a phenomenon which captures interest from a variety of disciplines, ranging from physics \cite{Jana2016} and biology \cite{Chaban2015} to bio-inspired design in engineering \cite{Yang2015}. 
Interest in motility is often in the context of chemotaxis, a behaviour where membrane bound proteins which act as chemo-receptors sense the presence of certain chemicals in the environment and affect the flagellar rotation in order to move towards or away from the source \cite{Wadhams2004}.
Bacteria in nature have evolved, and typically live, in complex environments like the mammalian gastrointestinal tract or the soil. However, the majority of studies have focused on the case of dilute aqueous media with a single chemical gradient  \cite{Wadhams2004, Sourjik2012}. While this reductionist approach has been invaluable and generated a large body of knowledge about underlying mechanisms of bacterial chemotaxis, our interest is now shifting towards understanding how robust is bacterial navigation when multiple competing stimuli are present \cite{Wadhams2004, Berg2017, Sourjik2012}.
Briefly, and taking the example of the model organism \textit{Escherichia coli}, the bacterium swims by rotating a flagellar filament bundle that propels its body through the environment \cite{Berg1973,TurnerBerg2000}. 
Each flagellum consists of a long thin helical filament attached to a bacterial flagellar motor (BFM), which drives its rotation at rates exceeding 100 Hz. It spins predominantly in the counter-clockwise (CCW) direction with occasional switches to clock-wise (CW) \cite{SowaBerry2008}. As long as all the filaments are spun CCW they form a stable bundle and when one or more participating flagella switches to CW rotation, unbundling occurs resulting in a so-called "tumble" event that brings a change in swimming direction once all the flagella resume CCW rotation \cite{TurnerBerg2000}.
In a homogeneous environment tumbles happen stochastically, whenever enough copies of the phospho-CheY protein \cite{Wadhams2004} (CheY-P) diffuse to the motor and increase the chance of a CCW-CW switch through their interaction with the BFM \cite{Sarkar2010,Bai2010}. As a result, a single bacterium moves in the pattern of a random walk \cite{Berg1972}. The intracellular fraction of CheY that is phosphorylated is controlled by transmembrane proteins which act as chemosensors \cite{Wadhams2004}. They are able to bind very specific chemicals in the cell exterior and, in response, transiently increase or decrease the concentration of CheY-P inside the cell. This provides a mechanism for biasing the random walk by making tumbles more or less probable and ensuring, for example, that tumbles are less probable if the cell is moving towards a source of food.
Only transiently modifying its reorientation probability allows the bacterium to quickly respond to new changes in its surroundings and navigate gradients rather than just have a binary response to presence or absence of a chemical \cite{Vladimirov2009}. This behaviour, where the sensors modulate their own sensitivity to bring the CheY-P concentration to baseline levels only seconds after responding to a stimulus, is called perfect adaptation \cite{Segall1986}. It, however, works only if the successive stimuli are in approximately the $\mu$M range and thus are not over saturating the sensors \cite{Mao2003}.

Biased random walk is not restricted to bacteria but is ubiquitous in the biosphere. Variations of it describe the movement patterns that arise when large herbivores search for new grazing patches \cite{Knegt2007} and the way drosophila larvae search for optimal environmental temperatures \cite{Luo2010}. Additionally, it also has promising use in bio-inspired swarm robotics as a target search strategy \cite{Yang2015} and means of controlling the spatial extent of the swarm \cite{Obute2019}. Since the changes in the rotational direction of single motors are at the basis of this kind of directed motility in bacteria, the quantity that is commonly used to express how often cells change direction is the CW Bias, the total amount of time the motor spent rotating CW in a given interval, divided by the duration of the interval\cite{Segall1986, Bai2010, Rosko2017}. Since most studies on \textit{E. coli}'s biased random walk have been done in dilute environments, where the steady state CheY phosphorylation activity is generally low, and consequently the CW Bias is low, the efficiency of biased random walked has not been explored from the point of view of steady-state CW Bias changes. 

Recently, there has been interest in scenarios without perfect adaptation, where the CW Bias does not return to its pre-stimulus levels \cite{Meir2010} . Furthermore, experiments at  higher osmolarities, at values typical for the gastrointestinal tract, have found long term changes in CW Bias following a shift in solute concentration \cite{Rosko2017}. Because the biased random walk arises from transient changes in the CW Bias due to the concentration gradient, and on the level of individual bacteria, changing the steady state CW Bias can affect motion of the bacterial population in a complex manner.

Here, therefore, we explore the effect of different steady state CW biases on the ability of bacteria, or robotic swimmers, to quickly and accurately find their target. 
To answer such questions requires us to look in detail at the behaviour of whole bacterial populations. For the purpose, we introduce a new path integral model of chemotaxis that sums over whole trajectories of individual bacteria. The model can easily be modified to efficiently take account of interactions between the bacteria and non-local time delay effects in the chemotactic response. We find that a change to a single cell phenotype (steady-state CW Bias) leads to changes in multiple aspects of the population. Hence finding an optimal value for it is non-trivial and context dependent.

\begin{figure}
\centerline{\includegraphics[width=0.9\linewidth]{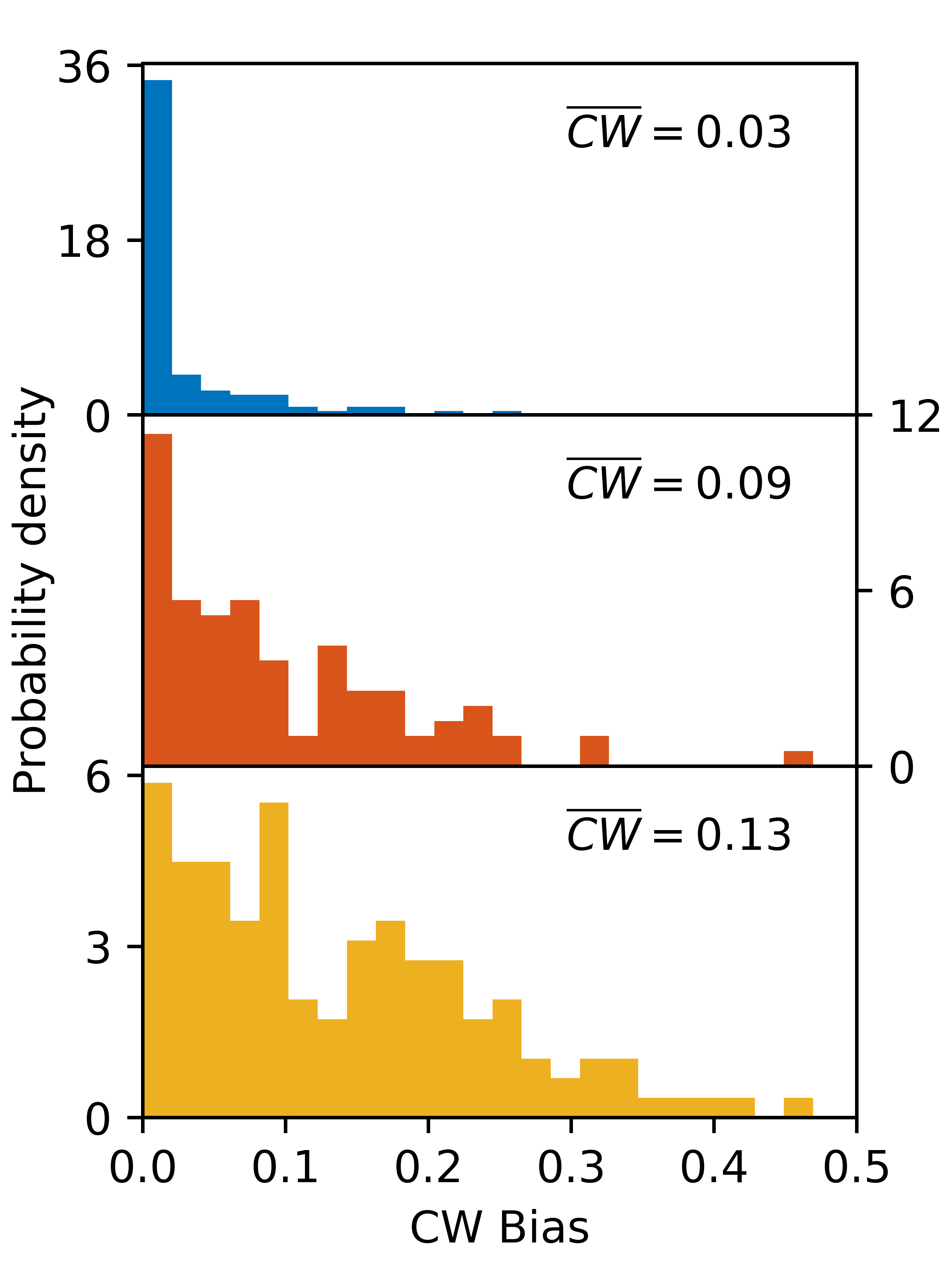}}
\caption{\label{Figure1} Variation in flagellar motor's CW Bias (A) CW bias histogram for cells in VRB Buffer (see \textit{Supplementary Information} for experimental setup and buffer composition). The data set includes a total of 118 single motor recordings, each 1~min long. (B) Histogram for cells in VRB Buffer with addition of 200~mM sucrose, containing 95 1~min recordings. (C) Histogram for cells in VRB Buffer with addition of 400~mM sucrose, containing 142 1~min recordings. Mean values of CW Bias are shown in each panel.}
\end{figure}

\begin{figure}
\centerline{\includegraphics[width=0.9\linewidth]{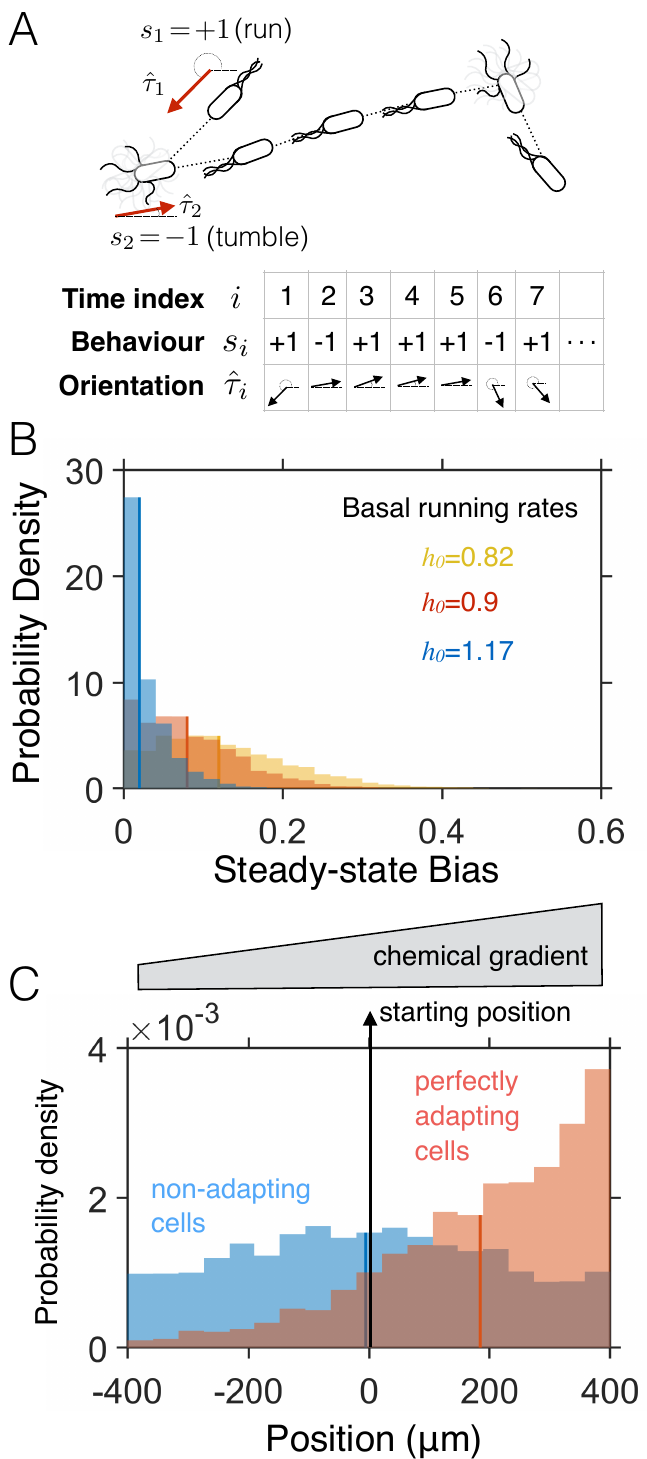}}
\caption{\label{Figure2} %\textcolor{red}{I am not sure we need the title 'Chain model of bacterial chemotaxis' in the figure? It can stay but I find it a bit distracting} 
Persistent trajectory  model of chemotaxis captures its basic features. (A) A chemotactic trajectory is modelled as a stochastic sequence of states, describing bacterial behaviour (either $s=+1$ for `run' or $s=-1$ for `tumble') and direction of movement ($\hat{\tau}$). (B) Distribution of steady-state tumbling bias (fraction of time spent tumbling) for different values of the basal running rate, $h_0$, fitted to match the mean CW Bias under different experimental conditions (see Fig.~\ref{Figure1}). Distribution of steady-state tumbling bias calculated from 10~s trajectories. Population size $N=10^5$ cells. (C) After 20~s, perfectly adapting cells ($B>0$) start to accumulate up a chemical gradient, whereas non-adapting cells ($B=0$) are incapable of performing chemotaxis. Population size $N=5000$ cells; chemical gradient 0.01 AU$\cdot\mu$m$^{-1}$.}
\end{figure}

We begin by extending our data set from \cite{Rosko2017}; we measure the CW Bias in 3 different media  (see \textit{Supplementary  Information}), showing a clear change in the CW Bias distribution and mean value, Fig.~\ref{Figure1}. Next, to study the impact of such profound changes in the steady-state CW Bias on the speed and accuracy of finding the target with bias random walk, we propose a statistical mechanical description of chemotaxis of individual bacteria, Fig.~\ref{Figure2}, which can be readily used as tool to quantify chemotactic behaviour in time-lapse movies of swimming bacteria.

We develop a parsimonious model of bacterial chemotaxis, composed of trajectories of single cells that are stochastic sequences of runs and tumbles. It has some mathematical similarities to  wormlike chain (WLC) models describing semi-flexible biopolymers \cite{Storm2003,Chakrabarti2005}.  
Its optimal length-scales are intermediate between those of PDE models \cite{Keller1970}, which capture average properties of populations but are insensitive to microscopic details, and agent-based models that link behaviour of individual bacteria to intracellular biochemistry \cite{Taylor2012,Neumann2014,Kwangmin2016} but are difficult to scale to experimentally realistic large populations. Interactions in space and time between the bacteria are also easy to implement in our framework. We explicitly calculate probability distributions of any function of chemotactic trajectories as a path integral. The path integral is defined as the weighted sum over all possible individual trajectories of the bacteria, thus the model naturally links the CW Bias of individual bacteria with the behaviour of the population. 
In this letter, the underlying biochemistry is included in a minimal way but it is easy to generalise the model to deal with more detail. 
 
We describe a chemotactic trajectory within a concentration field of chemoattractant/repellant, $c(x)$, as a chain of random steps (indexed by $i$). Each step $i$ represents the state of the bacterium over a time window $\Delta t_i=t_{i+1}-t_{i}$ in terms of (i) its chemotactic behaviour $s_i$, either `run' ($s_i=+1$) or `tumble' ($s_i=-1$); and (ii) its direction of movement, $\hat{\tau}_i$. Furthermore, bacteria move only when in the `run' state at speed $v$. Hence, the bacterial position, $x_i$, follows from the chain description of the chemotactic trajectory and the initial position $x_1$:
\[x_i = x_1 + v \sum_{j=1}^{i-1}  \hat{\tau}_j \dfrac{(1+s_j)}{2} \Delta  t\; .  \] 
State transitions, $i \rar i+1$, from $s_i,\hat\tau_i$ to $s_{i+1},\hat\tau_{i+1}$ depend only the states, $i,i+1$. Hence, starting in state $x_1$, the probability of observing the sequence $\left(s_1,\hat\tau_1,s_2,\hat\tau_2,\ldots ,s_T,\hat\tau_T\right) \equiv 
%[s(i),\hat\tau(i)]= 
\{s_{1:T}, \hat{\tau}_{1:T}\}$, where $T\gg 1$, is given by \[P(\{s_{1:T}, \hat{\tau}_{1:T}\}|x_1) = p_1 \prod_{i=1}^T p_{i \rar i+1}  
%(\{s_{i-1:i}, \hat{\tau}_{i-1:i}\}) 
\propto \exp\left(-H(\{s_{1:T}, \hat{\tau}_{1:T}\})\right) \] with weight defined by: 
\begin{eqnarray*}
H(\{s_{1:T}, \hat{t}_{1:T}\})  & = & {\epsilon \over 2} \sum_{i=2}^T (1-s_is_{i-1}) 
%\\ & & 
- \sum_{i=1}^T { h_i \over 2}  (1+s_i) \\ 
& & + \sum_{i=2}^T \kappa(s_{i-1}) (1 - \hat{\tau}_i \cdot \hat{\tau}_{i-1})
\end{eqnarray*}
Increasing $\epsilon$($>0$) in the first term of $H$ penalises  transitions  between `run' and `tumble' states, noting that a typical run could extend for several steps. In the second term, $h_i$ controls the preference for `running' over the `tumbling', which in general will  depend on the exposure of the bacterium to the chemoattractant/repellent as it moves through the concentration profile, $c(x)$. 
Here, we enforce perfect adaptation by making $h_i$ depend linearly on the concentration gradient, i.e.,  $h_i=h_0 + B \hat{\tau}_i \cdot \nabla c(x_i)$. Parameter $B$ controls the strength of the chemotactic response to the chemical gradient, with $B>0$ indicating chemoattractants and $B<0$ chemorepellents. Furthermore, $h_0$, hereafter referred to as the basal running rate, controls the distribution of steady-state tumbling bias (fraction of time a bacterium spends tumbling) (Fig.~\ref{Figure2}B).
%Perfect adaptation is encoded in the fact that $h(c(x))$ depends only on the gradient of the concentration, with $B$ quantifying its response to the gradient (Fig.~\ref{Figure2}C). 
For simplicity, and without the loss of generalization, we assume every change in rotational direction of a motor results in a tumble, hence hereafter we use the terms CW Bias and tumbling bias interchangeably. The model could be extended to include any mathematical relation between cell's run/tumble bias and the number and CW Bias of the motors, such as the one experimentally observed previously \cite{Mears2014}.  Finally, the third term of $H$  controls the change of orientation between steps, which depends on the chemotactic state. Since reorientation is significantly larger during tumbling, 
\[\kappa(s_i) = \left\{ \begin{array}{ll}
						\kappa_> & s_i=-1 \\
						\kappa_< & s_i=+1 \\
                                    \end{array} 
\right. \]
with $\kappa_> \gg \kappa_<$.  
The final bacterial position is given by $x(t_T) = x_j, j=T$.  

Setting $\epsilon=h=0$ and $\kappa$ constant reduces this model to the classic wormlike chain of polymer physics. We evaluate the path integral numerically, using a constant time step equal to the duration of a typical tumble event, i.e., $\Delta  t = 0.1$s \cite{Berg1978, Berg1972}; constant speed corresponding to the average running speed on glucose, i.e.,  $v=20\mu$m$\cdot$s$^{-1}$  \cite{Berg1972, Jana2016}; and $\epsilon=1$, $B=1$, $k_>=1$, $k_<=0.1$.
% We note that the model can be extended to handle non-constant time-steps as well as variable running speeds within the population.
 %(\textit{E. coli}'s velocity can change based on the metabolic state, but here we keep it constant with the speed corresponding to running on glucose (\textcolor{red}{cite Berg Brown 1972 and big paper Jana 1st author})). 
 
 %results

\begin{figure}
{\includegraphics[width=0.9\textwidth]{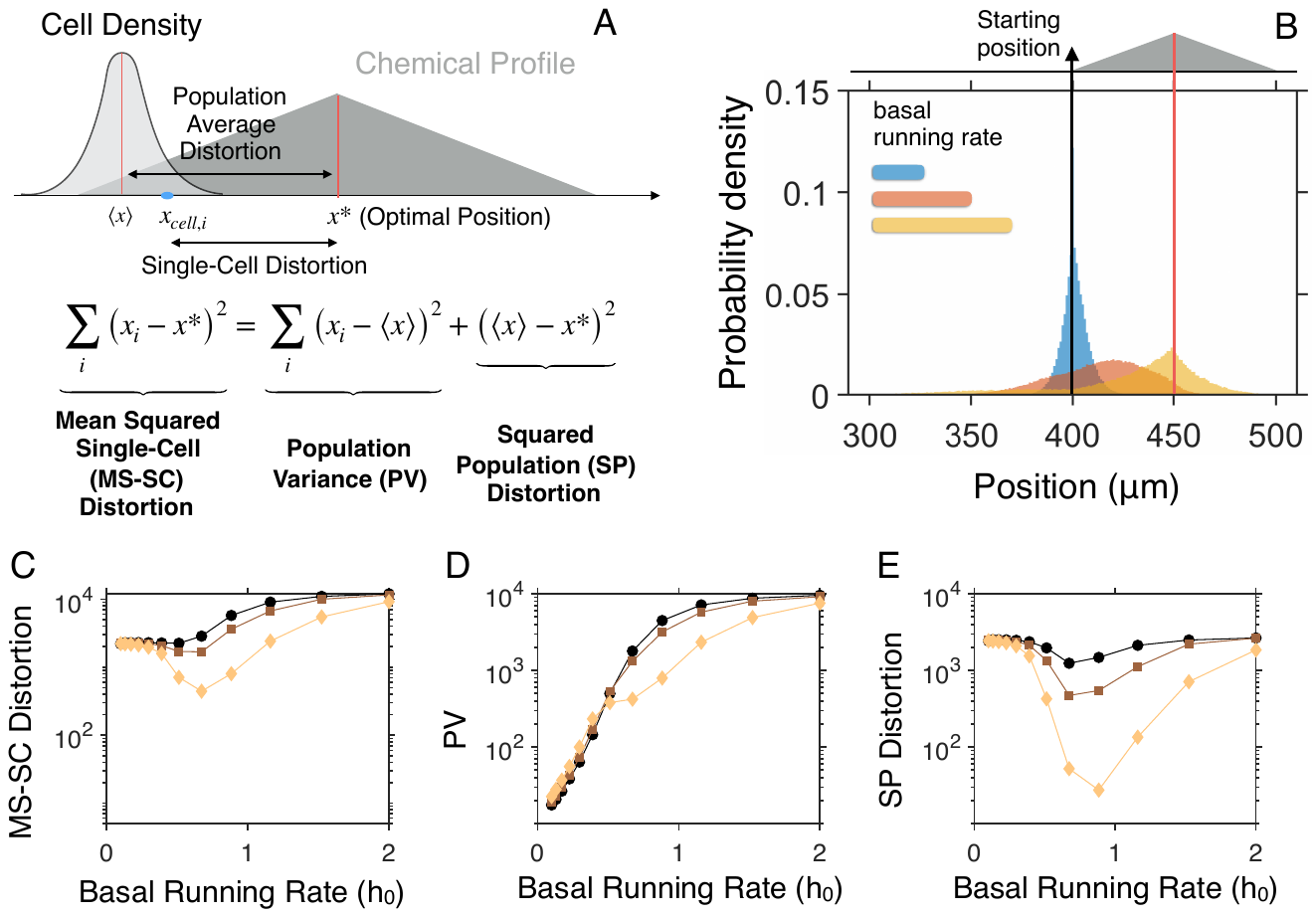}}
\caption{Bacterial chemotaxis speed and accuracy is influenced by the basal running rate. (A) Schematic illustration of a triangular chemotactic profile, in which chemotactic bacteria will seek to move towards the optimal position ($x^*$). We use the term distortion to denote the distance of single bacteria to $x^*$, and population-average distortion for the distance between the population average position and $x^*$. MS-SC distortion, PV and SP distortion stand for mean squared single-cell distortion, population variance and squared population distortion as defined in the main text. (B) Accumulation of bacterial populations with different basal running rates in a triangular chemical profile. Each population consists of $N=10^6$  cells, initialised at the left base point of the triangular profile and followed over 10s. (C) Mean squared single-cell distortion, (D) population variance and (E) squared population distortion as a function of the basal running rate for different heights of the triangular profile. Markers correspond to different gradients of the triangular profile, i.e., $0.005$ ($\bigcirc$), $0.01$ ($\Box$), and  $0.02$ ($\Diamond$) AU$\cdot\mu$m$^{-1}$   }\label{Figure3}
\end{figure}

To understand the effect of  steady state tumbling bias on the speed and accuracy of finding a target, we focus on finding the peak of a triangular-shaped profile of chemoattractant in one dimension, Fig.~\ref{Figure3}A. We follow the chemotactic response of bacterial populations, each with a different basal running rate ($h_0$), initially positioned at the tip of the base of the triangular profile. Fig.~\ref{Figure3}B illustrated how the basal running rate modulates the speed and accuracy with which cells find the target. Lower values of $h_0$ (Fig.~\ref{Figure3}B; blue population) achieve consistent exploration of the chemical profile and hence less cell-to-cell variability. However, this comes at a cost of a slower average movement of the bacterial population toward the target. As $h_0$ is increased a portion of the cells approach closer to the target, but the dispersion of the population increases, with a portion of cells completely missing the chemoattractant-rich area (left tail of the red population in Fig.~\ref{Figure3}B). Higher values of $h_0$ give rise to higher levels of heterogeneity, as prolonged running periods enables cells to disperse faster and miss the target (Fig.~\ref{Figure3}B, yellow population).  

To quantify these observations we introduce the mean squared single-cell distortion (MS-SC distortion), which is the mean squared distance of a single cell position ($x_i$) from the optimal position, i.e. the peak of the triangular profile ($x^*$):

\[
\sum_i \left(x_i-x^*\right)^2 = \sum_i \left(x_i - \langle x \rangle \right)^2 +  \left(\langle x \rangle-x^*\right)^2 
\] 
MS-SC distortion can be written as a sum of population variance (PV) and squared population distortion (SP distortion), where $\langle x \rangle = \sum_i x_i$ is the mean bacterial position. We note that the MS-SC distortion is always greater than the SP distortion as indicated by the variance decomposition formula.

Fig.~\ref{Figure3}C-E shows the three terms in the equation (MS-SC distortion, PV and SP distortion) as a function of the basal running rate ($h_0$), quantifying the connection between response accuracy and population heterogeneity.  For example, when faced with shallow gradients (Fig.~\ref{Figure3}C-E, black circles), single cells suffer on average from higher distortion as we increase the basal running rate, Fig.~\ref{Figure3}C. The effect  could go unobserved, if we fixate only on the mean of the population coming closer to the target (Fig.~\ref{Figure3}E) disregarding the fact that at the same time the variability in the population is increasing rapidly, Fig.~\ref{Figure3}D. Furthermore, for the range of gradients we examined, the SP distortion demonstrates a non-monotonic behaviour as a function of $h_0$. This suggests that the basal running rate could be  regulated to allow bacterial population to optimally adapt to different environments conditions.  The values of $h_0$ inferred from the CW Bias data (0.82-1.17; see Fig.~\ref{Figure2}) are close to the values achieving minimum SP distortion in Fig.~\ref{Figure3}. 

\begin{figure}
{\includegraphics[width=1\linewidth]{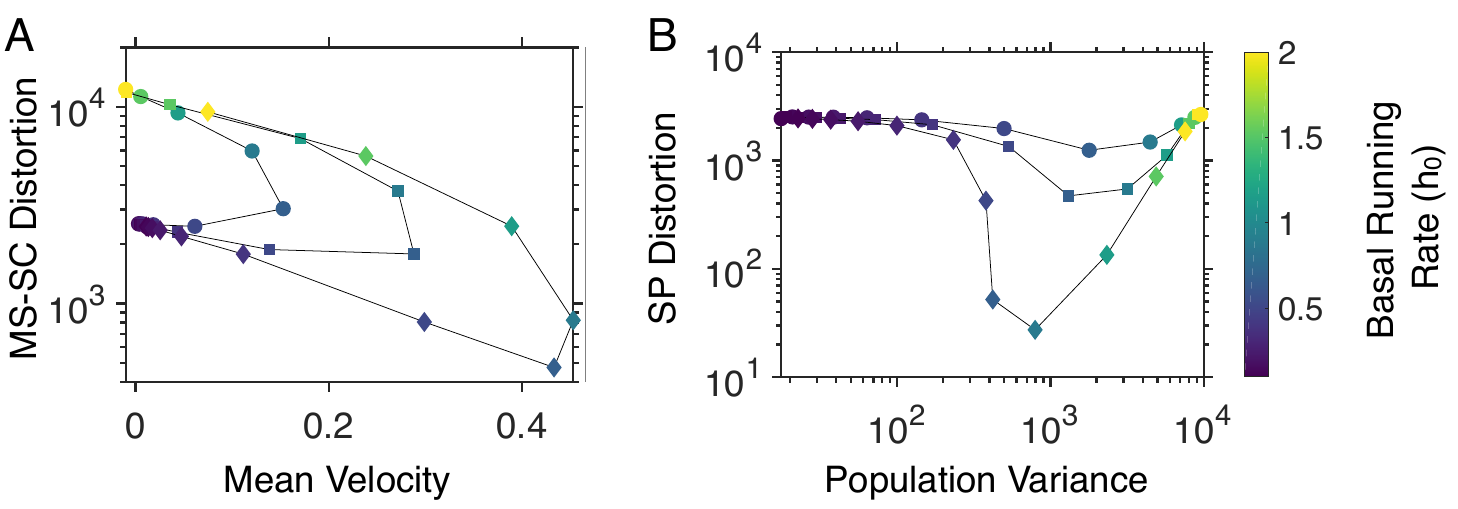}}
\caption{Trade off on chemotactic speed and accuracy imposed by different running rates. (A) Mean squared single-cell distortion versus mean velocity of a bacterial population in a triangular profile as the basal running rate (colour coded) is varied. Optimal basal running rate achieves the highest mean velocity and lowest distortion. Markers correspond to different gradients of the triangular profile ($0.005$ ($\bigcirc$), $0.01$ ($\Box$), and  $0.02$ ($\Diamond$) AU$\cdot\mu$m$^{-1}$). (B) Squared population distortion versus populational variance of a bacterial population in a triangular profile at different basal running rate (colour coded). The optimal basal running rate achieves the lowest distortion and positional variance simultaneously. Different lines correspond to different heights of the triangular profile. }\label{Figure4}
\end{figure}

Fig.~\ref{Figure4}A illustrates that basal running rate can be controlled to maximise chemotactic speed and accuracy. Low $h_0$ give rise to low population velocity due to the increased times spent in the tumble state. High $h_0$, on the other hand, enables cells to run for longer, but obstructs them from integrating adequate information about the chemoattractant concentration. The latter gives rise to higher MS-SC distortion as well as lower average velocity. Hence, for intermediate values of $h_0$ the system is able to demonstrate optimal levels of velocity and accuracy. Similarly, Fig.~\ref{Figure4}B  illustrates that successful chemotactic strategies for the entire bacterial population involve intermediate values of $h_0$, where the population variance remains low as the population mean comes close to the target (low population distrortion). We note that similarly  to the basal running rate, bacterial running speed also affects the speed and accuracy of bacterial accumulation and must be taken in account to optimise the chemotactic response in complex environments (see \textit{Supplementary Information}). 
%\textcolor{red}{really nice. Perhaps it is also worth discussing something about changes in swimming speed (and maybe tumble duration although that's less likely), so if the speed changes the ideal values of h0 could change too?}

Our model provides a novel, parsimonious description of bacterial chemotaxis at the single cell level, capturing all the key features of its phenomenology. The statistical character of the model provides access not only to single cell chemotactic dynamics as other agent-based chemotaxis models do \cite{Kwangmin2016}, but also allows computationally efficient estimation of population measures, bridging the gap between the two scales of description. Despite its simplicity the model can be straightforwardly extended to capture more realistic modes of chemtotaxis--involving, for example,  changes in the running speed of cells or not perfectly-adapting responses--and study how such modes affect bacterial accumulation. With advancements in observational techniques and manipulation methods used to probe bacterial chemotaxis in complex environments, our model could provide a useful inference tool for identifying tumble/run events and characterising single-cell chemotactic responses in more realistic scenarios. Finally, the influence of the reorientation frequency of individuals within a population, on the population level speed and accuracy of reaching a target could inspire search algorithms used in unmanned aerial vehicles \cite{Ha2018, Alfeo2018}.

\section*{Acknowledgments}
\noindent We thank members of Pilizota lab and Richard Berry, Filippo Menolascina and Marco Poiln for useful discussion. TP and JR acknowledge the support from the Office of Naval Research Global and Defense Advanced Research Projects Agency (GRANT12420502). TBL acknowledges support of BrisSynBio, a BBSRC/EPSRC Advanced Synthetic Biology Research Centre (grant number BB/L01386X/1). And TP, JR, TBL and MV acknowledge the support by the Grand Challenge NetworkPlus in Emergence and Physics Far From Equilibrium 2016-19, Engineering and Physical Sciences Research Council (EPSRC), reference EP/P007198/1. Lastly, we thank the organizers of workshop Physics and Biology of Active Systems held in June 2015, University of Aberdeen, where this collaboration started.

\newpage

\onecolumngrid

\twocolumngrid

\end{document}